# Pressure effect on superconductivity and magnetism in α-$FeSe_x$

L. Li,[1] Z. R. Yang,[2,*] M. Ge,[1] L. Pi,[1] J. T. Xu,[1] B. S. Wang,[2] Y. P. Sun,[2] and Y. H. Zhang[1]

[1]National High Magnetic Field Laboratory, University of Science and Technology of China, Hefei 230026, People's Republic of China

[2]Key Laboratory of Materials Physics, Institute of Solid State Physics, Chinese Academy of Sciences, Hefei 230031, People's Republic of China

In this paper, pressure effect on superconductivity and magnetism has been investigated in $FeSe_x$ (x = 0.80, 0.88). The magnetization curves display anomaly at $T_{s1}$ ~ 106 K and $T_{s2}$ ~ 78 K except for the superconducting diamagnetic transition around $T_c$ ~ 8 K. The magnetic anomaly at $T_{s1}$ and $T_{s2}$ can be related to a ferromagnetic and an antiferromagnetic phase transition, respectively, as revealed by specific heat measurements. The application of pressure not only raises $T_c$, but also increases both $T_{s1}$ and $T_{s2}$. This system shows clear evidence that superconductivity arises in a phase with strong magnetic character and the superconductivity coexists with magnetism. In addition, the specific heat anomaly associated with the superconducting transition seems to be absent.

---

*Corresponding author. Email address: zryang@issp.ac.cn

The discovery of superconductivity in doped LaFeAsO has generated much attention in layered FeAs systems [1-5]. The parent LaFeAsO material shows a structural transition from tetragonal to orthorhombic crystal symmetry at 150 K, followed by the formation of spin-density-wave (SDW) at a slightly lower temperature around 140 K [3]. Doping with fluoride suppresses the SDW and leads the onset of superconductivity. Therefore this system is proximity to magnetic instability, and the superconductivity in the doped systems seems to be of unconventional nature [4, 5].

Very recently, Hsu *et al.* reported the observation of superconductivity with critical temperature $T_c$ around 8 K in another Fe-based system, the PbO-type α-FeSe compound [6]. Subsequently, $T_c$ has been raised to 27 K with the application of high pressure [7]. Comparing with the layered FeAs systems, α-FeSe not only has the same planar sublattices but also displays structure and magnetism instability [8-10]. Density functional study showed that α-FeSe has the SDW ground state [11]. However, Lee *et al.* believed that the ground state for stoichiometric α-FeSe is nonmagnetic and the magnetism is driven by anion vacancy [12]. Upon cooling, a structural transition from tetragonal to triclinic symmetry around 105 K accompanied by magnetic anomaly was reported by Hsu *et al.* [6]. However, Margadonna *et al.* observed a tetragonal-orthorhombic structural transition at 75 K [13]. Although these results are inconsistent, α-FeSe seems to be a superconductor with strong magnetic character. Therefore a detailed investigation of the high temperature magnetism is needed.

In the paper, the superconductivity in PbO-type α-FeSe$_x$ has been examined with three nominal compositions (x = 0.80, 0.84 0.88). Especially, we investigated the pressure effect on superconductivity and magnetism for x = 0.80 and 0.88. All the samples shows superconductivity with $T_c$ ~ 8 K. With the decrease of temperature at ambient pressure, the field-cooling magnetization displays a sharp upturn around $T_{s1}$ ~ 106 K and an abrupt decrease at $T_{s2}$ ~ 78 K, in consistent with the results of Hsu *et al.* and Fang *et al.*. Specific heat measured for x = 0.88 shows that $T_{s1}$ and $T_{s2}$ can be related to a ferromagnetic and an anti-ferromagnetic phase transition, respectively. The applied pressure not only raises $T_c$, both also increases both $T_{s1}$ and $T_{s2}$, contrary to the conjecture of Margadonna et al.. In addition, the specific heat anomaly associated with the superconducting transition seems to be absent.

The polycrystalline samples with nominal compositions FeSe$_x$ (x = 0.80, 0.84, 0.88) were prepared by standard solid-state synthesis method [6]. High-purity powders of iron (99.9%) and selenium (99.9%) were mixed uniformly in a 2g batch, then sealed into evacuated quartz tubes and heat treated at 700 $^0$C for 24 hours. The initially sintered samples were ground and pressed into round-shaped pellets (10mm diameter, 2mm thick). The pellets were re-sealed in evacuated quartz tubes and sintered at 700 $^0$C for another 24 hours.

Structure and phase purity of the samples were examined by an x-ray power diffraction (XRD) method, with Cu K$_α$ radiation at room temperature. In consistent with the results of Hsu *et al.* [6], the prepared samples are composed of primarily PbO-type α-FeSe (*P4/nmm*) and tiny amount of impurity phases, the impurity was identified to be element Se and β-FeSe. The resistivity was measured using a standard four-probe method from 2.5 to 300 K in a Quantum Design Physical Properties Measurement System (PPMS). The specific heat was measured with thermal relaxation method in PPMS. Temperature dependence of magnetization was measured using a superconducting quantum interference device (SQUID) magnetometer. The application of pressure was performed in an Easylab Mcell 10 Pressure cell.

Figure 1 displays the temperature dependence of resistivity ρ in the field range from 0 to 5 T. Upon cooling, all the samples display metallic behavior before the onset of superconductivity around $T_c^{Res}$ ~ 10 K. With increasing magnetic field, the critical temperature decreases monotonously. By defining the critical temperature $T_c$ with criterion of $\rho_{cri} = 50\%\rho_n$, the upper critical field deduced at 0 K $H_{c2}(0)$ for x=0.88 is about 27 T, similar to the result of Mizuguchi et al. [7].

The temperature dependence of magnetization (M) was measured at 10 Oe in both field cooling (FC) and zero field cooling (ZFC) sequence for x = 0.80 and 0.88. Both samples at ambient pressure show clear superconducting diamagnetic response below the onset temperature $T_c^{mag}$ around 8 K, see Fig. 2. With increasing pressure, $T_c^{mag}$ increases. The estimated pressure coefficient d$T_c$/d$P$ is about 0.4 K/kbar, more than 10 times larger than that reported by Yeh et al. [9], but less than the value Mizuguchi et al. derived from the pressure effect on resistive transition [7]. Comparing with the results of Mizuguchi et al., the lower pressure coefficient might be due to different measuring method. See Fig. 1 and Fig. 2, the onset critical temperature of the resistive transition $T_c^{Res}$ is larger than that of the diamagnetic transition $T_c^{mag}$.

Figure 3 displays $M(T)$ curves at different pressures in a broad temperature range from 4.5 K to 200 K. At ambient pressure, both ZFC and FC magnetization shows anomaly at $T_{s1}$ ~ 106 K and $T_{s2}$ ~ 78 K, respectively. Upon cooling, the FC magnetization first increases abruptly at $T_{s1}$ then displays a sharp decrease and restores to its high temperature value at $T_{s2}$, signaling two magnetic transitions. Around the first transition temperature $T_{s1}$, Hsu et al. found a tetragonal-triclinic structure transition [6]. Near the second transition temperature $T_{s2}$, Margadonna et al. reported a tetragonal-orthorhombic structural transition, independently [13]. Therefore, the magnetic anomalies should be intrinsic behavior of the superconducting α-FeSe with tetragonal crystal symmetry, which is related to the structural and magnetic instabilities.

To understanding the nature of magnetic anomalies, we further performed specific heat measurement for x = 0.88 sample under magnetic field of 0 and 14 T, respectively. As shown in Fig. 4, the specific heat in zero field also displays anomalies around $T_{s1}$ and $T_{s2}$, indicting phase transitions at both temperatures. The applied magnetic field of 14 T increases $T_{s1}$, but depresses $T_{s2}$ to lower temperature. Therefore $T_{s1}$ and $T_{s2}$ can be related to a ferromagnetic and an antiferromagnetic phase transition, respectively. Around $T_{s2}$, the specific heat shows λ-like shape characteristic of a second order phase transition and confirms the bulk nature of the antiferromagnetict transition. Consistently, upon cooling the sample at 10 Oe to 4.5 K then warming back, no hysteresis of magnetization has been found around this temperature, see Fig. 5. The magnetization only displays hysteresis around $T_{s1}$. From the shape of the specific heat around $T_{s1}$, we believed that this temperature corresponds to a structural transition as observed by Hsu et al. Due to strong coupling between spin and lattice, the ferromagnetic transition is driven here. The antiferromagnetic transition at $T_{s2}$ might be caused by magnetic instability of the system.

For α-FeSe, the stoichiometric sample is nonmagnetic [12, 14], both magnetism and superconductivity are driven by anion vacancy. If there is competition between magnetism and superconductivity, the applied pressure should suppress the magnetic transition, as expected by Margadonna et al. [13]. However, contrary to their expectation, the applied pressure not only increases $T_c$ but also raises $T_{s1}$ and $T_{s2}$, see Fig. 3. Therefore, the current picture seems to be that superconductivity arises in a phase with strong magnetic character, and the superconductivity coexists with magnetism. More interesting, the specific heat anomaly associated with the superconducting transition in this material appears to be absent. See the inset of Fig. 4, $T^2$ dependence of $C_p/T$ displays a traditional linear behavior for metal. The applied field of 14 T also has no evident influence on the specific heat around $T_c$. At first glance, the absence of specific heat anomaly might imply non-bulk superconductivity. However, we do observe clear diamagnetic response. Therefore, another possibility may be related to its unconventional pairing mechanism, as proposed by Karchev et al for ferromagnetic superconductor [15]. Experimentally, the absence of specific heat anomaly has also been found in other unconventional superconductor, like the ferromagnetic superconductor $ZrZn_2$ and organic superconductors [16, 17].

In conclusion, the pressure effect on magnetism and superconductivity has been studied in $FeSe_x$ (x = 0.80, 0.84, 0.88). The magnetization and specific heat measurements show two magnetic phase transitions at $T_{s1}$ and $T_{s2}$, respectively. The application of high pressure not only raises the superconducting critical temperature, but also intensifies the magnetic transitions. This system shows clear evidence that superconductivity arises in a phase with strong magnetic character and the superconductivity coexists with magnetism. In addition, the specific heat anomaly associated with the superconducting transition appears to be absent.

**Acknowledgements:**

This research was financially supported by the National Key Basic Research of China Grant 2007CB925001, 2007CB925003 and the National Nature Science Foundation of China Grant 10774147.

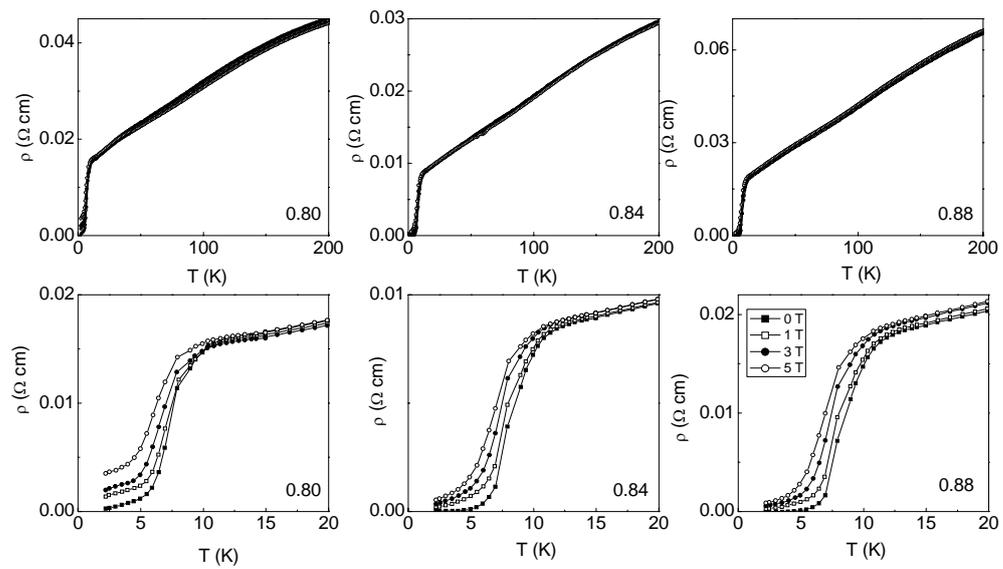

**Figure 1** Temperature dependence of resistivity for FeSe$_x$ (x = 0.80, 0.84, 0.88) at different fields.

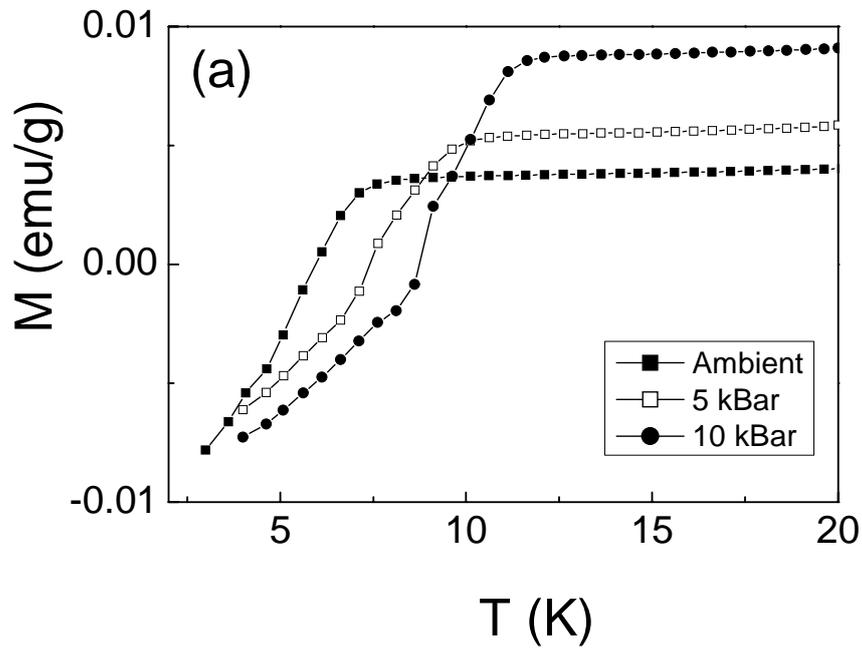

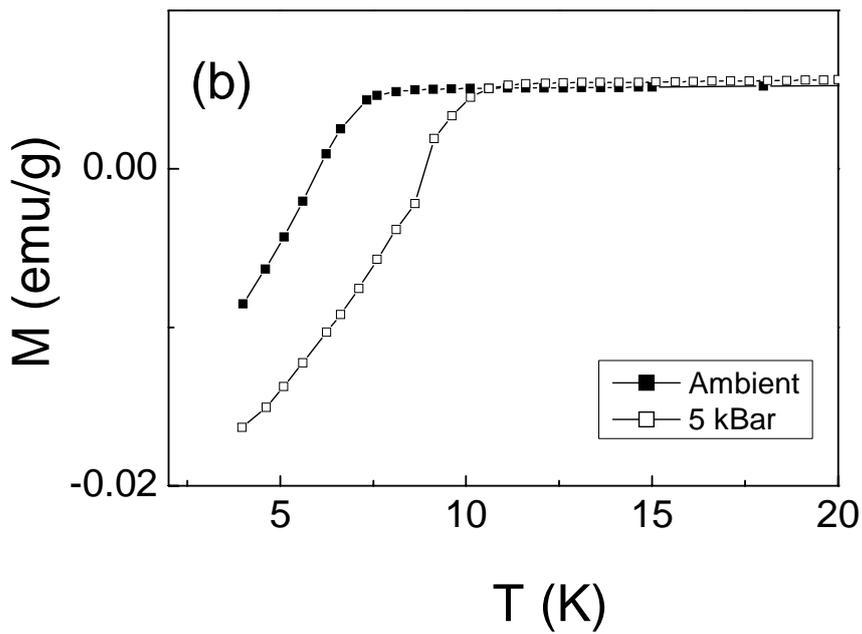

**Figure 2** Zero field cooling (ZFC) magnetization under different pressures around the superconducting transition temperature for (a) x = 0.80 and (b). x = 0.88.

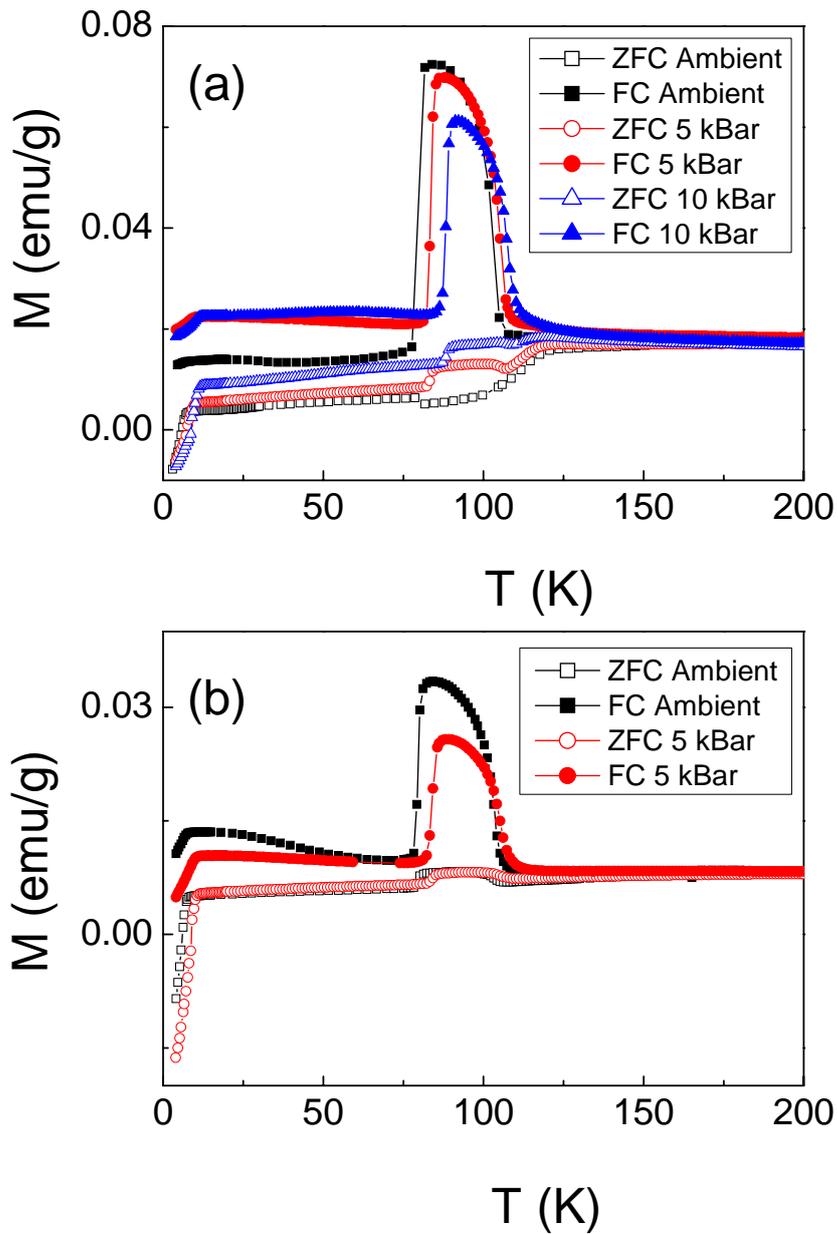

**Figure 3** Temperature dependence of magnetization under different pressures for (a). x = 0.80 and (b). x =0.88. The magnetization has been measured at 10 Oe in both field cooling (FC) and zero field cooling (ZFC) sequence.

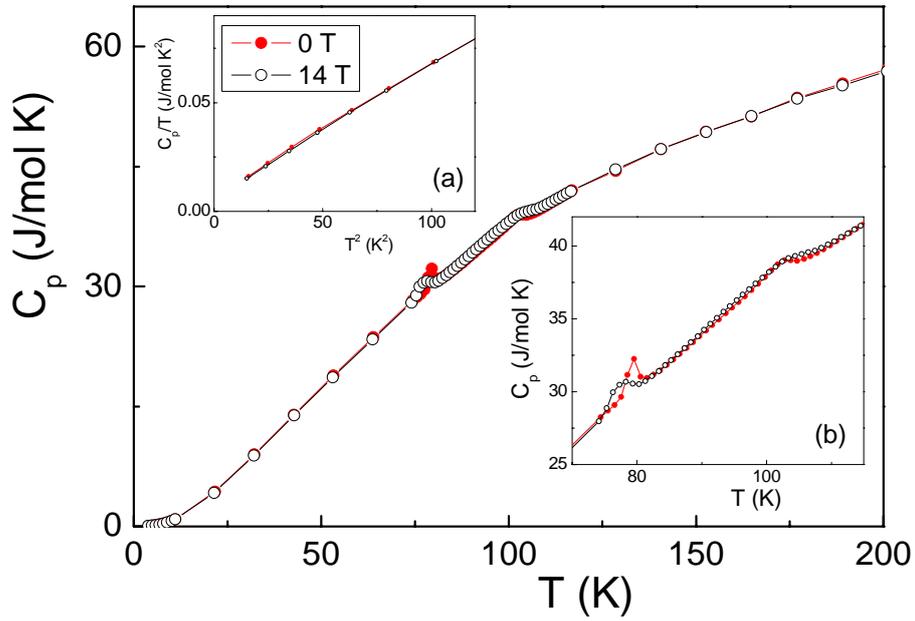

**Figure 4** Temperature dependence of specific heat for x = 0.88 sample under magnetic field of 0 and 14 T. Inset (a) displays $C_p/T$ as a function of $T^2$ around $T_c$. Inset (b) shows enlarge view of $C_p(T)$ around $T_{s1}$ and $T_{s2}$.

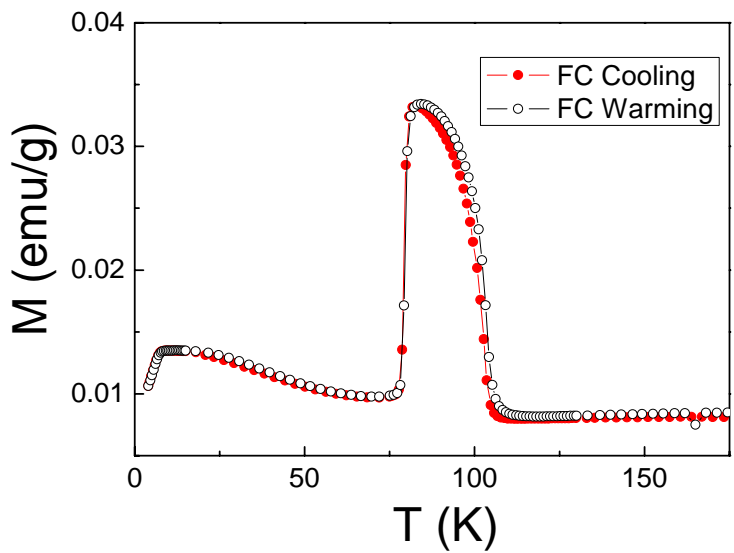

**Figure 5** FC magnetization curves with solid circles denoting field cooling and open circles denoting field warming sequence.